\documentclass[aps,prd,twocolumn,groupedaddress,nofootinbib,longbibliography]{revtex4-1}

\usepackage{graphicx}
\usepackage{amsmath,amssymb,amsfonts, amsthm}
\usepackage[colorlinks,citecolor=blue,linkcolor=blue,urlcolor=blue, breaklinks=true]{hyperref}
\usepackage{breakurl}

\usepackage{physics} 
\usepackage[percent]{overpic} 
\usepackage{float} 

\usepackage[english]{babel}

\usepackage{comment} 

\usepackage{xcolor}

\newcommand{\be}{\begin{equation}}
\newcommand{\ee}{\end{equation}}

\usepackage[autostyle, english = american]{csquotes}
\MakeOuterQuote{"}

\begin{document}

\title{A response to the Muci\~no-Okon-Sudarsky's\\ Assessment of Relational Quantum Mechanics}

\author{Carlo Rovelli}
\email{rovelli.carlo@gmail.com}
\affiliation{Aix Marseille University, Universit\'e de Toulon, CNRS, CPT, 13288 Marseille, France.}
\affiliation{Perimeter Institute, 31 Caroline Street North, Waterloo, Ontario, Canada, N2L 2Y5.}
\affiliation{The Rotman Institute of Philosophy, 1151 Richmond St.~N London, Ontario, Canada, N6A 5B7.}

\begin{abstract}
\noindent A recent paper by Muci\~no, Okon and Sudarsky attempts an assessment of the Relational Interpretation of quantum mechanics. The paper presupposes assumptions that are precisely those questioned in the Relational Interpretation, thus undermining the value of the assessment. 
\end{abstract}

\date{\small\today}

\maketitle

\section{The problem of quantum physics}

The problem of quantum physics is not that we have \emph{no} way of making sense of it.  The problem is that we have \emph{many} ways of making sense of it. But each of these comes with a high conceptual price.  

Each interpretation of quantum mechanics demand us to accept conceptual steps that for many are hard to digest.  Pilot-Wave like interpretations require a non-local layer of reality which is inaccessible in principle; Many-Worlds interpretations require zillions of real actual copies of ourselves seeing slightly different worlds; Qbism forces us to a strong instrumentalism; Physical Collapse models require physical processes that have never been observed; Relational Quantum Mechanics assumes that  contingent properties are sparse and relative. And so on. 

Richard Feynman observed that Nature often admits different interpretations of the same phenomena. He suggested that a good scientists should better keep  all alternative in mind, not knowing which one will next turn out to be good. Perhaps this is a good attitude for quantum theory. As science develop, one perspective may well turn out to be more fruitful. One price will turn out to be worth paying. 

In one form or the other, all approaches are radical (because the novelty of quantum theory to be radical), but the conceptual assumptions of the different interpretations are in general radically different. For this reason, communication between scientists and philosophers working within different interpretations is sometimes uneasy: it often reduces to an empty restatement of alternatives "beliefs".  

This is unfortunately the case of the "assessment" of Relational Quantum Mechanics (RQM) \cite{LaudisaFedericoandRovelli} recently attempted by Muci\~no, Okon and Sudarsky (MOS) \cite{Mucino}.   These authors are defenders of Physical Collapse models \cite{Okon2014, Okon, 
Bedingham2016, Sudarsky2021}. They give for granted a number of conceptual assumptions that ground the Physical Collapse interpretation of quantum mechanics. They evaluate RQM starting from the conceptual assumptions on which their own work is based, and find that that RQM does not solve \emph{their} problems. 

MOS  fail to see that RQM is coherent (as other interpretations are), but at the price of giving up a priori assumptions that they are not ready to give up.  Hence their "assessment" of RQM end up being empty: they do find contradictions, but the contradictions are not within RQM; they are contradiction between the RQM and assumptions that are given for granted by the author's, but are not commonly accepted.  

In this paper, I discuss in detail the assessment of RQM that they give, and explain in detail why it misses the point. 

\section{RQM in a nutshel}

This is not the place for a detailed description of RQM, for which I refer to the literature \cite{Rovelli:1995fv, LaudisaFedericoandRovelli, Rovelli2017b, DiBiagio2020}.  But for the purpose of clarity, I give here a compact account of this interpretation (as I understand it today).

The basic idea is that the world can be decomposed (in many alternative manners) into "physical systems" that interact among themselves.  Each physical system can be characterized by a set (in fact, an algebra) of physical variables $A_1,..., A_n $.  These variables do not represent how the system \emph{is}, but rather what the system \emph{does} to another system, when there is an interaction. Outside this context, they are not determined \cite{Calosi2020,Dorato}. 

Physical variables (i) take value only at interactions and (ii) the value  they take is relative to the interacting system.  The occurrence that a variable takes a value at an interaction is called a quantum event.  

For a given system $S$, its quantum theory specifies its variables $A_n$, the set of values $a_n$ that each of these can take (that is, their spectrum), and gives probabilities $P(a_n;b_n)$ for values $a_n$ \emph{relative to a second system $O$}, as a function of values $b_n$ \emph{relative to this same system $O$}.  

Values taken by variables are labelled by the systems in interaction, but it can be shown \cite{DiBiagio2020} that the effect of decoherence is to render the labelling irrelevant as soon as quantum interference is suppressed.

Key features of this interpretation, important for what follows are:\\
(a) The interpretation is realist in the sense that it describes the world as a collection of real systems interacting via discrete relative quantum events. \\
(b) The wave function is interpreted epistemically, in the same manner in which its classical limit, namely the Hamilton-Jacobi function, is. \\
(c) There are no special systems playing the role of "observers", no special role given to "agents", or "subjects of knowledge", no fundamental role given to special "measurement" contexts. \\
(d) The traditional tension between unitary evolution and wave function collapse is resolved by relativising values. That is, the evolution of the probability distribution of the values of variables \emph{relative to a system $O$} is predicted by unitary evolution only as long as $S$ is isolated and does not interact with $O$.  This does not conflict with the fact that variables of subsystems of $S$ can take value with respect to one another, because the quantum mechanical transition amplitudes only connect values relative to the same system. In a slogan: with respect to Schr\"odinger's cat the poison is definitely out  or not, but this has no bearing on the possibility of an external observer to observe quantum interference effects between the two alternatives.   \\
(e)   A coherent picture of the world is provided by all the values of variables with respect to \emph{any single} system;  juxtaposing values relative to 
\emph{different} systems generates apparent incongruences, which are harmless because they refer to a non-existing  "view from outside the world".\\ 
These minimal notes should be sufficient to address the points raised by MOS.

\section{How MOS define the problem of quantum mechanics}

The characterisation of the problem of quantum mechanics (QM) that MOS give is idiosyncratic. It makes sense if we remember that they expect that the only way to solve it is to introduce a physical collapse mechanism.  According to MOS, the problem of QM is indeed the following: say a Stern-Gerlach apparatus entangles the position of a particle with the $z$ component of the spin. Schematically, with obvious notation: 
\begin{equation}
|\psi\rangle\sim |1\rangle \otimes |+\rangle_z+|2\rangle \otimes |-\rangle_z.
\ee
This same state can also be written as 
\begin{eqnarray}
|\psi\rangle &\sim& \Big(|1\rangle+ |2\rangle\Big) \otimes |+\rangle_x+ \Big(|1\rangle-|2\rangle\Big) \otimes |-\rangle_x.
\end{eqnarray}
The problem of QM, according to MOS is: why then this same apparatus does not measure the $x$ component of the spin?  In their words:  "Without information not contained in a complete quantum description, standard quantum mechanics is unable to deliver concrete predictions regarding the possible final outcomes of an experiment."  

This is a bizzarre characterization of the problem. Of course standard quantum mechanics is able to deliver concrete predictions regarding the possible final outcomes of an experiment, as any experimenter can testify.  The reason is that "standard quantum mechanics" is not \emph{only} a quantum state evolving unitarily.   Standard quantum mechanics as formulated in textbooks (mostly in the Copenhagen language) includes much more: operators associated to measuring procedures, eigenvalues, the existence Bohr's macroscopic world and so on.  Once you fold these ingredients in, predictions are clear.   

Of course there are good reason to consider the Copenhagen interpretation unsatisfactory, and interpretations that have a larger applicability arguably exist. With the possible exception of Many Words, in general other interpretations do \emph{not} start by simply discarding everything except the evolving state. Hence the characterization of the problem given by MOS is, at most, a potential objection to Many Worlds, certainly not the general formulation of the problem of quantum theory. 

What equations (1) and (2) show is that there are (special) cases in which the bi-orthogonal decomposition of a state is not unique. This is well known. It is a potential problem only for the interpretations of quantum mechanics that rely on the bi-orthogonal decomposition, such as modal interpretations \cite{Kochen1985, Dieks1988,Dieks1994} or some versions of Many World.  The bi-orthogonal decomposition theorem plays no role in many other interpretations and in particular in RQM, which is about variables, not about states. 

More explicitly, this is the definition of the problem of QM given my MOS: "The formalism [of RQM] ends up critically depending upon the notion of measurement---which is a problem because such a notion is never precisely defined within the theory. And it is not only that the standard theory does not specify when a measurement happens, it also does not prescribe what it is that is being measured (i.e., in which basis will the collapse occur)."  

Notice that the problem is \emph{formulated} in terms of "measurement".  But the notion of "measurement" plays no role in RQM.  In fact, the entire logic of RQM is to give up any notion of "special" interactions that should count as "measurements".

Since MOS work in Collapse Models, their assumption is that that reality is described objectively and universally by an evolving wave function. This wave function evolves unitarily until something "special" happen. This special is the "measurement". Recall that in Collapse Models something (a fundamental frequency \cite{Ghirardi1986a}, a threshold in the gravitational self-potential \cite{Penrose1996a},...) must determine the occurrence of the physical collapse. Under this logic, MOS asks whether there is something "special" that determines when a measurements happen in RQM. The answer is that there is no "measurement" in RQM.   There is no universal objective wave function either, in RQM. 

The formulation of the problem of QM according to MOS is predicated on the basis of assumptions that are explicitly rejected in RQM. 

\section{Which variable takes value}

Nevertheless, we can still try to translate the question posed by MOS into the language of RQM and see if it may refers to anything. The question could be which variable of a system takes value relative to another system (that is, when does a quantum event happen), and when does it do so.  

Recall that the setting of RQM is not a uniformly evolving quantum state.  It is a setting in which two specific distinguishable physical systems are singled out, say $S$ and $P$. Quantum mechanics gives descriptions of the world conditional to this (arbitrary) choice and describes how one system affects the other \emph{when they interact}. The theory can therefore be applied anytime we have two well distinct systems interacting. 

Which variable takes value in the interaction is dictated by the physics: in the classical theory, we can describe the interaction between the two systems, say, in terms of an interaction term in the Hamiltonian that depends, in particular, say, on a variable $A$ of the system $S$: then $A$ is the value that takes value. The reason is that the interaction Hamiltonian depends on the property of $S$ responsable in determining the effect of $S$ on $O$. And this is precisely how quantum theory describes the world (in RQM): the way systems affects one another. 

The formulation of the problem in terms of the states (1) and (2) above does not even make sense in RQM: the theory is about values of variables, not about states. 

\section{When does a variable take value}

To the question of the \emph{time} when the quantum event happens, the answer is similar. It happens when the systems interact. In turn, we may ask when do the systems interact.   The answer is the (quantum) physics of $S$, given its dynamics and the interaction terms in its hamiltonian.   For instance, suppose that $S$ is an electron in a radioactive atom and $P$ is a Geiger counter.  When does a quantum event relative to these two systems happen?  

Notice that the Schr\"odinger wave function of the electron leaks \emph{continuously} out of the atom and is therefore constantly in causal contact with the Geiger counter. The electric force of the electron on the Geiger counter, irrespectively of the position of the electron is never exactly zero.  If you start with a wave function ontology you have the problem of understanding what happens when the Geiger counter clicks and when it does so.  But the RQM ontology is not the continuous wave function leakage. It is the actual quantum event of the clicking of the Geiger counter, which is discrete. When does it happen?  Knowing the Hamiltonian of the system and the interaction hamiltonian with the affected system, standard quantum mechanics can be used to determine the probability distribution in time of the occurrence of the interaction. That is, quantum events are discrete, their occurrence can be predicted only probabilistically,  and the probability distribution of their occurrence can be computed using quantum mechanics itself and the specific quantum dynamics in play. 

The difficult conceptual step here is to accept the idea that continuity is a large scale approximation, while the happenings of the world are discrete and probabilistic at the quantum scale.  In its relational interpretation, quantum theory describes a world that is fundamentally discrete and probabilistic. Exchanges between system are always discrete and regulated by the quantum of action $\hbar$.

This is precisely the original intuition of Max Born and its collaborators in G\"ottigen \cite{Born1925,Born1926,Fedak2009,Capellmann2020}, who were the first creators of quantum theory. But it is also the conclusion to which their opponent, one of the most strenuous defenders of continuity and deepest thinkers about the problems of the theory, Erwin Schr\"odinger, ended up accepting:  "There was a moment when the creators of wave mechanics (that is, himself) harboured the illusion of having eliminated discontinuity from quantum theory. But the discontinuities eliminated from the equations of the theory reappear the moment the theory is confronted with what we observe. [...] it is better to consider a particle not as a permanent entity but rather as an instantaneous event. Sometimes these events form chains that give the illusion of being permanent, but only in particular circumstances."\cite{Schrodinger1996}

MOS reject this logic a priori, because they interpret quantum measurements as "special events" that happen to a realistically interpreted wave function under peculiar circumstances to be specified, not as the general happening of all phenomena. 

Within MOS's Physical Collapse logic, RQM does not answer MOS's question. But this is not because of an incoherence of RQM, it is only because MOS have a priori assumptions that are rejected in RQM. The assumptions of RQM are similar to the original ones of Born and collaborators. The assumptions of MOS are the early ones (lately rejected) by Erwin Schr\"odinger. 

\section{When does measurement happen}

In their quest for finding a rule for an objective and system independent condition for quantum measurement, MOS make reference to an old article with a latin title "Incertus tempus, incertisque loci: when does a measurement happen" \cite{Rovelli1998c}.  What MOS search, namely an objective criterium singling out quantum measurement, is not in that paper, because the paper addresses a different problem.  

In RQM terms, that paper asks the following question.  Suppose that two systems $S$ and $O$ interact among themselves.  If we consider \emph{only} values of variables relative to a third system, $P$, is there anything we can say \emph{in terms of these values} about the timing of a quantum event realised by the interactions between $S$ and $O$?

At first sights the answer seems to be negative, because the values of the variables with respect to $P$ are blind to the values of the variables of $S$ with respect to $O$. Yet, \cite{Rovelli1998c} points out that there is an operational sense that can be given to this question. The intuition is that if Wigner has a friend making a measurement in a closed box, Wigner --with sufficient knowledge of what is in the box-- can say something about \emph{when} his friend makes the measurement, even if he has no access to it.    This intuition is made concrete in \cite{Rovelli1998c} as follows.  The previous interactions between the $S\cup O$ system and $P$ give probabilistic predictions about what would $P$ see (that is, how $S\cup O$ would affect $P$) if an interaction with $P$ happened at some arbitrary time. Now consider an interaction where $O$'s "pointer" variable affects ("is seen by") $P$.  The predictions can include the timing when the pointer move, even if they do not include which direction it will moves.  Hence $P$ can predict "when the measurement happen" under a special definition: with respect to $P$, a measurement between $O$ and $S$ happens when $P$ can predict that the the pointer variable has moved.  Since the prediction is probabilistic, this gives only a probability distribution of course. That is, figuratively, "I say that the measurement has happened with probability $p$ if I predict that I will find the pointer moved (correlated with the $S$'s variable, which $P$ can independently detect) with probability $p$".   This peculiar operational definition does in fact correspond to what one would concretely say in a laboratory, but has nothing to do with the foundation of RQM.   MOS misinterpret it as foundational.   

MOS observe that this operational definition is based on the notion of quantum event itself.  This is correct, or course. Their objection is that this implies a  regression at infinity.  This is wrong, because the notion of quantum event does not require such indirect operational definition. The observation counts as an objection only if one asks, as MOS do, an absolute non-relational determination of quantum events, and searches it in  \cite{Rovelli1998c}. But a non-relational definition of quantum events is exactly what RQM mechanics demands \emph{not} to ask, in order to make sense of quantum physics.

\section{Observer dependence}

An explicit criticisms of MOS regards a statement contained in the paper that has inspired RQM, the 1995 paper \cite{Rovelli:1995fv}. The paper says that ``the experimental evidence at the basis of quantum mechanics forces us to accept that "distinct observers give different descriptions of the same events".  MOS object that this not true and cite pilot-wave theory as contrary evidence. Here MOS are confusing two different facts. One is the assumption realised in the pilot-wave theory that there exist a universal objective state of affairs. A different one is the account that a real observer can give of a set of events.  Since the pilot-wave theory is a hidden variable theory, observers do not have access to the universal objective state of affairs. This is hidden.  In particular, the global wave function postulated by the pilot wave theory is in principle inaccessible to observers (otherwise the pilot wave theory could be used to make non-probabilistic predictions and beat standard quantum theory, which is not the case). In fact, what happens in the pilot wave theory is that observers can make predictions using "effective" quantum states.  These are relative states and observer dependent.  Thus confirming the observation in \cite{Rovelli:1995fv}. Therefore the MOS objection has no ground. 

The other argument that MOS cite as evidence against the statement above is --not surprisingly-- physical collapse model.  It is of course true that collapse model do violate the observation in  \cite{Rovelli:1995fv}, but they violate the predictions of standard quantum theory as well (which is obvious by the fact that they are  empirically distinguishable from standard quantum theory), while \cite{Rovelli:1995fv} states clearly that everything it says is within the validity of standard quantum theory. RQM makes sense of quantum theory (the most successful physical theory ever) as it is, not under the assumption that it fails. This can be seen directly: the detailed argument used in  \cite{Rovelli:1995fv} to derive the above statement is based on the existence of certain quantum interference effects predicted by quantum theory and not predicted by Physical Collapse  models. 

In the course of this discussion, MOS reiterate their prejudice about quantum collapse ("In sum, a core conceptual problem of standard quantum mechanics has to do with an ambiguity regarding \emph{the dynamics}."), they interpret observer dependence as due to the fact that quantum collapse is disregarded ("Such an ambiguity allows for different observers to give different descriptions of the same events.") and essentially blame RQM only for not adhering to their view. 

\section{Self measurement}

A long section of the paper is devoted by MOS to criticise an observation about an assumed impossibility of self measurement in \cite{Rovelli:1995fv}. The observations about self measurement in that old paper are indeed vague. In fact these vague considerations have been abandoned in later presentations of RQM and replaced by sharper definitions \cite{LaudisaFedericoandRovelli}. For instance, in the condensed summary given above, variables are interpreted as describing the way a system affect \emph{other} systems.  The point is of course that the "R" in RQM stands for "relations": RQM is an account of quantum physics in terms of relations, namely relative variables. It assumes that physics is about relative variables, describing how systems manifest themselves to other systems. 

MOS clearly misunderstand this. For instance they write "Returning to Rovelli's statement that there is no meaning in being correlated with oneself, we must say that we find such a claim quite odd. [...] it is clear that the different parts of an observer certainly are correlated between them: her left hand is never more that 2 meters away from the right one."  This objections betrays the misunderstanding that MOS have of RQM. In the case considered, of course one hand has a position with respect to another hand. The two hands can interact, exchange light signals, etcetera. But it is so precisely because they are two!  What is the position of a single hand if there is noting else with respect to which can be defined?   Position is in fact a quintessential example of a relational property in physics: it is always defined with respect to something else.

\section{Decoherence and unitarity}

Another misunderstanding is in MOS's reading of the paper \cite{DiBiagio2020} which explore the role of decoherence in RQM.  MOS write "More recently Rovelli [...] has argued that decoherence plays a crucial role in explaining the breakdown of unitarity". This is a misunderstanding of the paper. Decoherence plays no role in the breaking of unitarity.  The unitary evolution of the probability distribution the values of the variables of the system $S$ with respect to a system $P$ holds only as long as $S$ can be considered isolated. At every interaction of $S$ with something else, unitary evolution of $S$ alone breaks down simply because $S$ is not anymore isolated.  If $S$ interacts with a second system $O$, the unitary evolution of $S\cup O$ (relative to $P$) stills hold, but when $S$ interacts with $P$ there is definitely a breaking of the unitary evolution of its variables \emph{with respect to $P$}.  Unitary evolution holds for \emph{any} system, but only as long as it refers to probabilites of interactions with something external. 

All this has nothing to do with decoherence.   The role of decoherence in RQM is completely different. How is it possible that a stable world described by variables that are \emph{not} relative emerge form the quantum world where variables are always relative? The paper \cite{DiBiagio2020} analyses this question and shows that decoherence, which is a concrete physical phenomenon, perfectly accounts for this. 

MOS argue in detail (and correctly) that decoherence alone does not solve the problem of the interpretation of quantum mechanics; they refer to their own Physical Collapse prejudice that the problem of quantum mechanics is to find a concrete mechanism that breaks unitarity;  they misinterpret \cite{DiBiagio2020} as an argument to say that decoherence is the source of the breaking of unitarity of RQM; and thus conclude, mistakenly, that there is a problem.  Breaking of unitarity is a trivial consequence of the perspectival aspects of RQM.  That is: evolution of \emph{anything} is unitary, as long as it refers to how this "anything" affects something considered external to it. 

\section{Ontology}

There is an interesting observation in MOS, regarding the quantum event ontology.  They observe that a quantum event given by the manifestation of the position of a particle can be precisely located in spacetime, but not so for example for the momentum.   This fact was noted earlier, in fact it was clear to Max Born himself \cite{Capellmann2020} in his 1925 formulation of quantum theory \cite{Born1925,Born1926}, and is emphasized in Niels Bohr long ruminations about complementarity between position and momentum.  This is general in quantum theory: since sharp momentum implies spread position, attributing a sharp momentum to a particle is not a statement regarding a sharp spacetime location. 

I think that the impossibility to always sharply locate  values of variables is just an indirect aspect of the non locality of quantum theory.  RQM does not aspire to get rid of all non local aspects of quantum theory, which are just aspects of Nature.   The Schr\"odinger equation of two particles describes physics that in a very precise sense is non-local.  Nature has certain some local aspects, but we should not impose to Nature the prejudice that everything works as in relativistic \emph{classical} field theory. 

Another objection in MOS is that RQM leads to solipsism or "quasi-solipsism" (whatever this means). This in fact is a common objection to RQM in popular online forums, but is a funny one for a perspective entirely based on relations!   It is true that RQM assumes that variables take value relative to systems, but nothing in RQM prevents the world to include complex "observers" like humans, that can converse among themselves and compare what they "observe" as observers.  In fact, the theory itself guarantees that if two such observers measure (this time the term is appropriate) the same quantity and compare notes, they find agreement. This was discussed in great detail in \cite{Rovelli:1995fv} and was crucial to show that RQM is not incoherent \cite{Fraassen:2010fk,pittphilsci3506}, as was sometimes feared in its early days.

There is an aspect of the ontology that is left open in the RQM literature: whether a system ontology or a quantum event ontology is more convincing. I think that RQM is compatible with both, and each has its own appeal. In the first, we assume that what exists is an electron, and its manifestations are sparse quantum event.  In the second, we assume that what exists are sparse quantum events and we call electron their ensemble and their dynamical relations. I think that both views are viable. After all, we can say that there is a chair, and its manifestation are all the perspectives on it, or we can say with Hume, that a chair is nothing else that the coherent ensemble of all its manifestations.  I see no reasons for which QM should decide a metaphysical issue that is open independently from quantum theory itself.

A similar remark hold for the formulation of the theory as a principle theory in terms of information. This perspective was emphasized in \cite{Rovelli:1995fv}, and has had a determined influence on the later development of diverse informational interpretations of quantum physics, including Qbism \cite{Fuchs2001}.  These ideas have lead to important successes in the reconstruction of quantum theory from information theoretical axioms \cite{Hoehn2014,Hohn2017,Hohn2017a}.  But the risk of excessive emphasis on the language of information and the ambiguity of the word "information" which is used with wide variations of meaning risks to push towards what seems to me an excessive instrumentalist stance. If there is one thing I agree with MOS is that what we wish to get from quantum theory is a credible account of, as far as we know, the world works.

\section{EPR}

The paper \cite{Smerlak:2006gi} analysis the EPR scenario at the light of RQM.  MOS object to that analysis posing the following  problem: The two spatially separated observers measure spin "along the same axis, at space-like separation from each other. Suppose, moreover, that, in accordance with the possibilities allowed by RQM, they both obtain spin-up. After performing such a measurement, both of them travel to the mid-point between their labs and, at the same time, announce their results. What would then happen?"   MOS argue that according to RQM a contradiction may emerge because nothing prevents A and B from reporting the same spin, against angular momentum conservation. This is a factual mistake.  What "happens" in RQM is how a system affect another system.  Without specifying with respect to which system are the variables taking value, the question is meaningless. MOS want to avoid referring to either A or B as observers.  This is possible in RQM.  It suffice to ask how A and B affect an external system, call it $P$.  The probability that A and B report contradictory results can be calculated using standard quantum mechanics, and is zero. The mistake of MOS is the common source of confusion in the EPR scenario, according to RQM: forgetting that A and B are themselves quantum systems and treating them classically.  If we assume that they are classical, we generate the contradiction. If we keep their quantum nature in mind, there is no contradiction in any set of events relative to a single system.  It is to juxtapose events relative to different systems that creates the illusion of a conflict. Facts are genuinely relative \cite{Frauchiger2018a,Bong2019,Brukner2018,Waaijer2019}, a point of view "from outside the universe" is genuinely excluded \cite{Dorato2013a}.

\section{Locality}

The discussion of locality in MOS reflects the general pattern. MOS take locality as the requirement that there are local beable in the sense of Bell that obey Bell local causality.  They then argue that his is not the case in RQM.  This is true in fact, and is precisely what is argued in the paper \cite{Martin-Dussaud2019a}. What is then argued in \cite{Martin-Dussaud2019a} that in the light of RQM this non locality is a simple consequence of the fact that some variables are not defined at some times.   MOS equivocate on this statement, reinterpreting it as meaning that the variables are well defined but not deterministically determined. No surprise on this, because MOS assume that there is always a unique determined state, this is part of their assumptions, assumption abandoned in RQM.  Hence they misunderstand \cite{Martin-Dussaud2019a} and wrongly conclude that the conclusion of \cite{Martin-Dussaud2019a} is "false". 
\vspace{.6cm}

\section{Conclusion}

Altogether, the paper  \cite{Mucino} turns out to be an exercise in misunderstanding. This is not surprising, because the paper is really an attempt to make sense of RQM in the context of the hypotheses underlying a Physical Collapse interpretation of Quantum Theory.  It is an attempt of interpreting RQM on the basis of assumptions rejected by RQM. 

The article could be read as an indirect way to promote the Physical Collapse perspective by using RQM as a straw man. It is written as if it was based on an "objective" position, based on commonly accepted assumptions. But the facts and assumptions that it gives for granted are far from being commonly accepted. They are the specific assumptions of the authors' "camp".  The authors have employed this strategy elsewhere \cite{Berjon2020a}. This invalidates the value of the assessment. 

Unfortunately, this happens repeatedly in debates about the interpretation of quantum mechanics. Defenders of one interpretation construct detailed arguments trying to show that another interpretation is "wrong", failing to see that what they are doing is simply applying their own hypotheses to the logic of authors that have alternative hypotheses. 

The difficulty of quantum mechanics is not about carefully articulating consequences of a shared conceptual framework: it is to understand which conceptual framework is more promising for making sense of the theory.  As mentioned in the introduction, every interpretation of QM includes conceptual steps hard to digest.  The exercise of criticising the details of an interpretation without accepting these conceptual steps  is futile.  I am sure we can do better, in articulating a fruitful conversation between different ways of making sense of quantum theory. \\

\centerline{***}
\vspace{4mm}

I thank the authors of \cite{Mucino} for having given me access to a first version of their paper before posting it publicly. This reply has been largely constructed on this first version.  This work was made possible through the support of the QISS grant ID\# 60609 of the John Templeton Foundation. 


%

\end{document}